# Internet of Things Search Engine: Concepts, Classification, and Open Issues

Tran Khoi Nguyen, *University of Adelaide*
Quan Z. Sheng, *Macquarie University*
M. Ali Babar, *University of Adelaide*
Lina Yao, *University of New South Wales*
Wei Emma Zhang, *Macquarie University*
Schahram Dustdar, *TU Wien*

**Abstract:** This article focuses on the complicated yet still relatively immature area of the Internet of Things Search Engines (IoTSE). It introduces related concepts of IoTSE and a model called meta-path to describe and classify IoTSE systems based on their functionality. Based on these concepts, we have organized the research and development efforts on IoTSE into eight groups and presented the representative works in each group. The concepts and ideas presented in this article are generated from an extensive structured study on over 200 works spanning over one decade of IoTSE research and development.

## 1 Introduction

Advancements under the moniker of the *Internet of Things (IoT)* allow things to network and become the primary producers of data in the Internet [1]. IoT makes the state and interactions of real-world available to Web applications and information systems with minimal latency and complexity [2]. By enabling massive telemetry and individual addressing of "things", the IoT offers three prominent benefits: *(i)* spatial and temporal traceability of individual real-world objects for thief prevention, counterfeit product detection and food safety via accessing their pedigree; *(ii)* enabling ambient data collection and analytics for optimizing crop planning, enabling telemedicine and assisted living; and *(iii)* supporting real-time reactive systems such as smart building, automatic logistics and self-driving, networked cars [3]. Realizing these benefits requires the ability to discover and resolve queries for contents in the IoT. Offering these abilities is the responsibility of a class of software system called the *Internet of Things search engine (IoTSE).*

IoTSE is a complicated and relatively immature research topic. The diversity of its solution space is, arguably, a primary challenge hindering its advance. Such diversity manifests itself in terms of *(i)* the type of operations within an IoTSE instance (e.g., discover content, index, and resolve queries), and *(ii)* the types of IoT content on which those operations are applied. Each combination of operation and content type represents a research area within the IoTSE literature with its own set of technical, social, and political issues. For instance, the IoTSE instances that discover and resolve queries on real-time sensing data from IoT-enabled sensors face the challenge of ensuring the "freshness" of data used for processing queries while minimizing the costly operation of pulling the data from sensors. IoTSE instances working with the actuating functionalities of IoT-enabled things, on the other hand, concern more with understanding the semantics of these functionalities. Due to the diversity of the IoTSE solution space and the lack of a shared vision of what IoTSE is and what it does, it is challenging to communicate the problems and the solutions related to this system. The lack of such models and constructs for the communication of IoTSE inhibits more extensive research and development efforts that span research communities over an extended time, which are necessary for the advance of the IoTSE. As the existing studies on IoTSE have primarily focused only on technical issues related to a particular "IoTSE operation – IoT content type" combination, and as the existing reviews and surveys on IoTSE have primarily focused on a particular type of IoTSE, the lack of models and constructs to communicate and classify IoTSE, which are applicable to its diverse solution space, has not been addressed in the existing literature.

In this article, we introduce the fundamental concepts related to the functionality of an IoTSE instance and a model called meta-path to provide a comprehensive yet succinct description of IoTSE instances by their functionality. We report a classification of IoTSE instances based on their meta-path description and present the representative IoTSE prototypes in each class. Finally, we discuss several open issues in the IoTSE research and development.

## 2 Methodology

The concepts and models presented in this article were generated from a structured and comprehensive study of the existing research works and industrial projects falling under the moniker of IoTSE. Our methodology was inspired by the systematic literature review method [4]. It comprised four phases: detection, selection, extraction, and synthesis (Figure 1). The *Detection phase* involves identifying potentially relevant articles from various academic sources. The *Selection phase* involves selecting a subset of articles that were high-quality and relevant to the study. The *Extraction phase* involves extracting raw data relevant to the questions of the study. The *Synthesis phase* involves synthesizing raw data into knowledge to answer questions of the study.

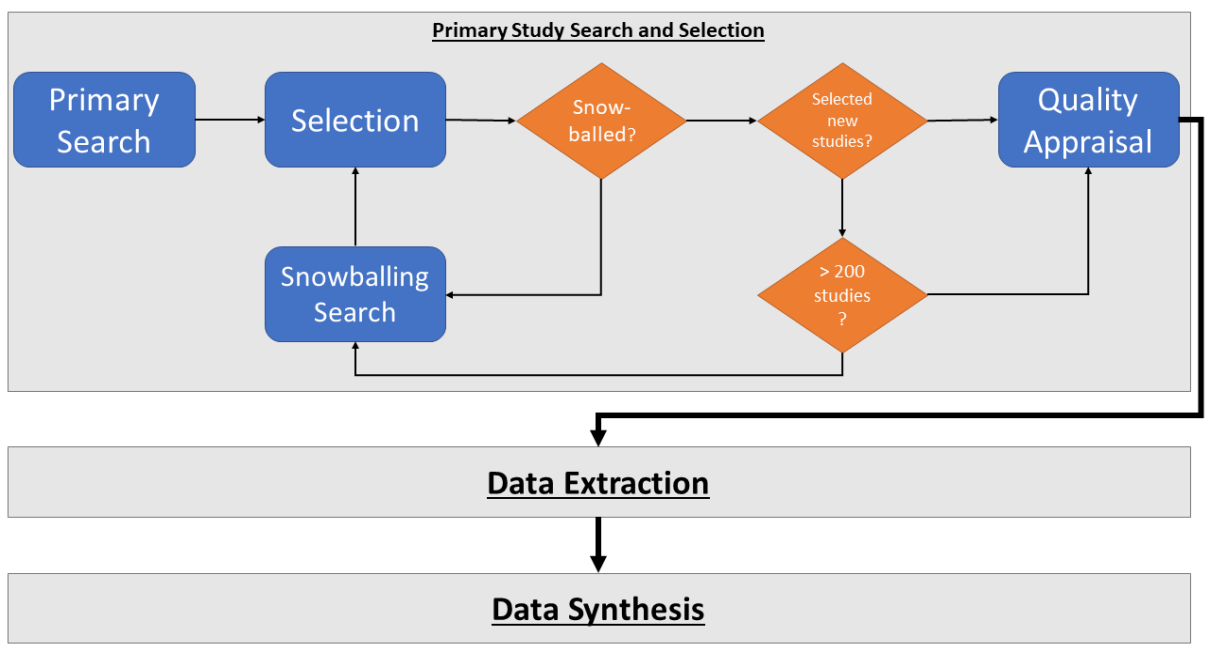

*Figure 1. Research Methodology*

Our method deviated from the systematic review method by using software tools for automation. Particularly, the detection and selection phase employed an in-house developed tool that queried academic data sources (i.e., *"primary search"*) and retrieved articles that had been referenced by the articles detected in the primary search (i.e., "snowballing"). We performed the primary search on various academic data sources, including the XML dataset of DBLP, with the Boolean query "search OR discover AND internet of things OR web of things". We assessed articles that emerged from the primary and snowballing search against the following selection criteria:

- Excluding papers that focus exclusively on physical and network layer.
- Excluding papers that focus on utilizing the sensing data from the IoT to extend the Web search
- Excluding the information retrieval papers that do not address IoT, *unless they are highly referenced by other relevant works*.

In the extraction phase, we extracted from papers the conceptualization, functionality, and internal operations of the reported IoTSE prototypes. Finally, we synthesized the extracted data into the concepts and models reported in this article.

By applying the reported method, we identified over 200 relevant works on IoTSE that span over a decade. Figure 2a compares the changes in the number of IoTSE-related works published and referenced between 2001 and 2016. The number of IoTSE works published each year has been increasing steadily since 2001. From 2009 – the birth year of the IoT, this number has risen sharply and peaked at 38 works a year in 2014 and 2015. There was a drop in the number of published works in 2016, which we contribute to the fact that our primary study selection concluded by the end of that year and therefore missed the accepted-yet-unpublished works.

The changes in the number of referenced IoTSE works have not assumed a similar pattern with the number of published works, however. Between 2001 and 2010, most of the published works were referenced at least once by other works. However, over the following six years, this number dropped gradually, and the gap between the number of published and referenced works widened. By 2015, only 13% of published IoTSE works received in-field citations.

Figure 2b compares the number of referenced IoTSE works and the number of in-field citations between 2001 and 2016 to provide more insights into the distribution of *attention* among the IoTSE literature. Despite fluctuations, the number of in-field citations rose steadily from 2001. After peaking at 67 citations in 2010, this number began to drop sharply. From these figures, we can see that a group of 29 works published between 2010 and 2012 received over 45% of the total number of in-field citation. This result might indicate that the perception of what an IoTSE instance is and what it should do have been driven by a subset of IoTSE works. A comprehensive analysis on the IoTSE literature and the internal operations of representative IoTSE prototypes is available elsewhere [5].

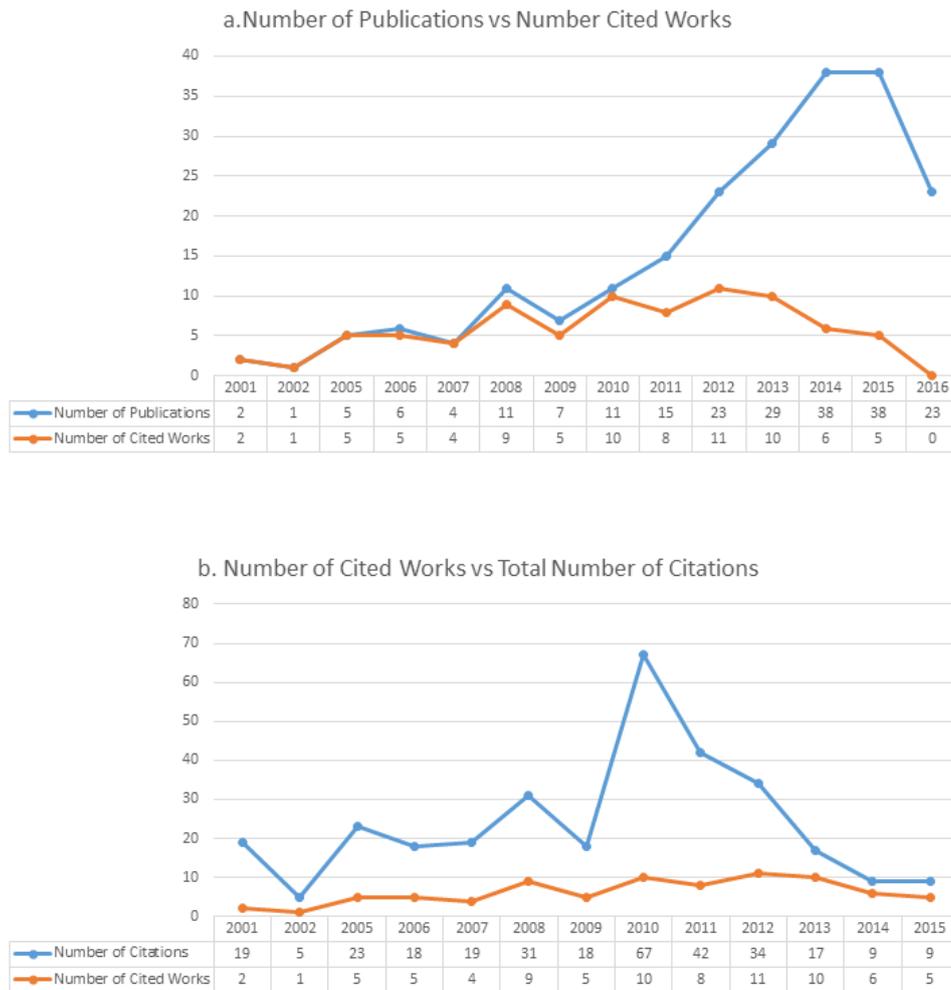

*Figure 2. Statistics regarding publication count and number of in-field citations of collected articles.*

## 3   IoTSE Concepts

### 3.1   Internet of Things Content

The Internet of Things comprises *IoT things* – physical objects enhanced with computing and networking capabilities and are potentially accessible via the Internet. For instance, a light bulb equipped with microcontrollers and wireless communication capability is an IoT thing that is commonly found in home automation applications. IoT things offer *IoT content*, such as the digital representation, data records, real-time sensor readings, and functionality, that are offered by or related to things.

The IoT content appearing in the IoTSE literature can be organized into four types: *representation, static information, dynamic information, functionality*. Figure 3 depicts four IoT content types of an IoT-enabled lightbulb. The representative content of the lightbulb comprises an HTML document that acts as a homepage of the light bulb for interacting with human users, and a JSON document that described the light bulb to machine agents. The dynamic information content of the light bulb denotes either the whole stream of energy consumption readings of the light bulb or the latest value in that stream. Due to the constant update of the light bulb, these contents are "dynamic". The static information content comprises the archived sensing data, the Web articles related to the light bulb, and the records of its journey across supply chains. Finally, the functionality content includes

*actuating services* that the lightbulb offers to alter its operation (toggling its power, changing its light color).

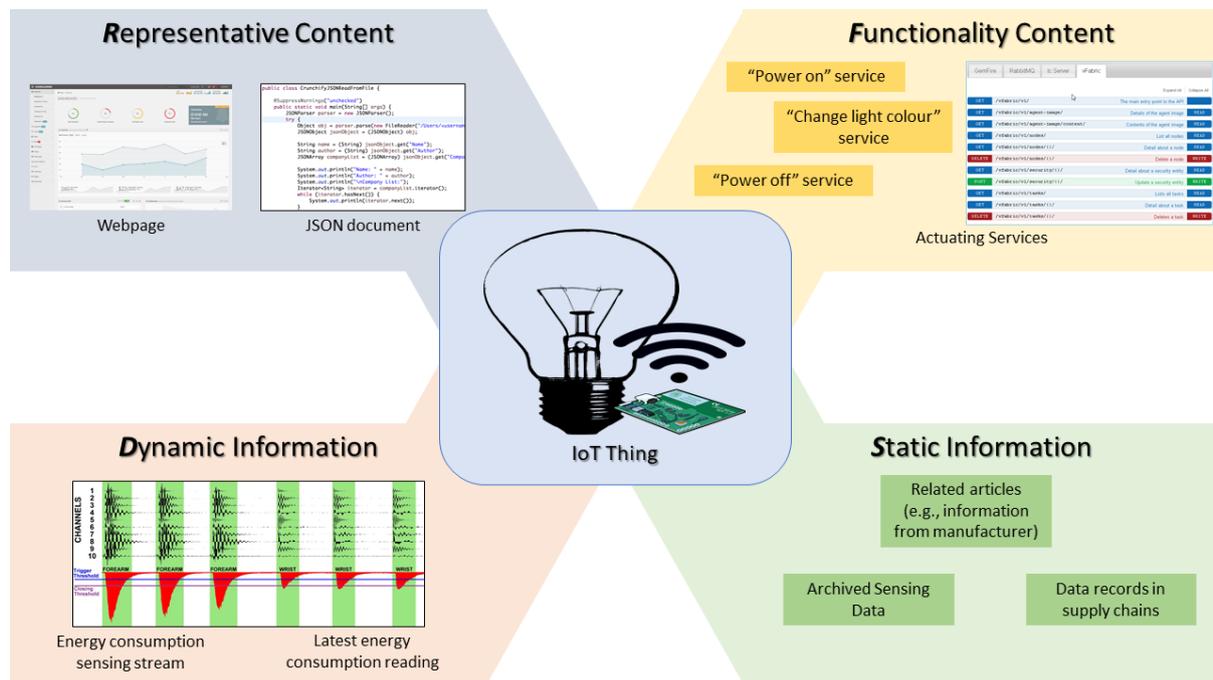

*Figure 3. Four types of IoT content of an IoT-enabled lightbulb.*

## 3.2 Discovery Activity

An IoTSE instance processes queries on various collections of IoT content. When these collections are not available, an IoTSE instance must carry out the *discovery activity* to detect IoT content in a local or global scope, and optionally collect the content into its internal storage. More than 90% of the assessed IoTSE prototypes include discovery activities.

On a global scale, the content discovery problem can be framed as a Web crawling problem to identify a subset of Websites that serve IoT content (i.e., IoT data sources) and retrieve the URI of the IoT content from those sites. In the existing literature, this crawling either relies on human's guidance [6] or the standard compliance of data sources [7]. On a local scale, the content discovery can be addressed as a wireless discovery problem, in which an IoTSE instance either broadcasts beacon signals for things to register themselves, or detects and queries things directly to retrieve their content [8, 9]. Local discovery can also be addressed as a service discovery problem in local area networks, using technologies such as multicast Domain Name System (mDNS) or Bonjour[1]. Semantic discovery is an alternative perspective on the content discovery problem. It concerns with detecting the semantics of IoT content and can be addressed by translation content to known data models [10].

## 3.3 Search Activity

The *Search activity* denotes the process of identifying a subset of discovered IoT content as search results of a given query. All assessed IoTSE prototypes covered this activity.

Formally, let $c$ be an item of IoT content, and $C$ be the collection of all content discovered by an IoTSE instance. For each query $q$, a set of contents $c_q$ that are relevant to the query exists. The task of an IoTSE instance is to construct the result set $\sim c_q$ that approximates the unknown $c_q$ by evaluating the relevance of each IoT resource against the given query with a relevance function $f(c,q)$. If the relevance function produces binary result, the process is considered *selection* or *lookup*:

$$Selection: c_q = \{c \in C | f(c,q) = 1\}, where\ f(c,q): C \times Q \to \{0,1\}$$

---

[1] https://developer.apple.com/bonjour/

If the relevance function produces a real value, the result set contains resources whose scores are higher than a predefined threshold $\alpha$. This process is called *resource scoring*.

$$Scoring: r_q = \{c \in C | f(c,q) > \alpha\}, where\ f(c,q): C \times Q \to \mathbb{R}$$

The selection process cannot determine the degree of relevance of IoT content, and therefore can be considered less advanced compared to the scoring process. However, we discovered that nearly half of the analyzed IoTSE prototypes utilized selection. Most of the remaining prototypes scored IoT content based on its distance from a given query in a multi-dimensional space.

The storage and indexing of the discovered IoT content link the discovery activity with the search activity. Most of the existing IoTSE prototypes address the heterogeneity of IoT content by limiting the type and format of content and handle each type independently. For example, IoT-SVK [11] utilizes two B+ trees and an R tree index to address textual description, numeric sensing data, and location of things separately. Some IoTSE prototypes, such as DiscoWoT [10], address the heterogeneity problem by mapping various formats of IoT content onto a common format for processing.

### 3.4 Meta-path

The lack of a descriptive and comprehensive model to communicate and classify the functionality of IoTSE instances was a major problem identified from our analysis. For instance, the term "object search" has been used to describe various types of IoTSE instances, which process queries on various types of content – real-time state, description, functionality of things – and return various types of IoT content including sensing data, location, data records, and actuating services of relevant IoT things.

The existing models describe an IoTSE instance either by the type of IoT content that is utilized for processing queries or returned as search results, without considering the relationship between them. Different from the previous types of search engine systems, such as Web search engines, IoTSE can utilize a combination of different IoT content types to assess a query and to derive search results. Moreover, the types of IoT content appearing in a query influence the internal operations of an IoTSE instance [5]. As a result, an IoTSE model must capture succinctly both the types of involving IoT content and the relationships among those types. Terms such as "object search" are inadequate. To address this issue, we propose a model called *Meta-path*.

Before defining meta-path, it would be helpful to introduce the idea of modeling the Internet of Things as a heterogeneous graph, which was inspired by PathSim [12]. The nodes in this graph consist of IoT contents and IoT things that own these contents. The edges that link things and content denote a possessive relationship between them. The edges that link things denote their possible correlations, such as sharing owners or operation environments [13, 27].

From a concrete graph, we can derive a meta-graph which presents relationships between types of nodes. Each node in a meta-graph is either a type of IoT content or a thing. An edge between a content type and a thing represents that the content type is offered by the thing. An edge between two things represents a correlation between them. Different types of thing-thing edges represent different forms of correlation between things. These thing-thing relationships can enable interesting queries such as finding all employees who had been in the meeting rooms that reported an abnormal energy consumption. However, we have not discovered IoTSE prototypes taking advantage of these correlations. Therefore, for the sake of clarity, we will not depict detailed thing-thing correlations in the following discussions.

Figure 4 depicts the IoT infrastructure in a smart building as a heterogeneous graph (upper) and the derived meta-graph. IoT things in this illustration consist of a smart light bulb, a meeting room, and a staff member who uses these facilities. The light bulb has seven IoT content items of four classes. The meta-graph captures relationships between IoT content types and things, as well as among things. For instance, two representative content items of the lightbulb are captured in the meta-graph as a single link between the representative content type and a thing.

A *Meta-path* is a sequence of edges on the meta-graph from one type of IoT content, through various IoT things, to another type of IoT content. In the IoTSE context, each meta-path can model the relationship between a type of IoT content used for assessing query and a type of IoT content used for deriving search results. By aggregating multiple meta-paths, we can model an IoTSE instances that utilize multiple types of IoT content.

A meta-path can be represented as follows:

$$(Query\ content\ type) \rightarrow Things * \rightarrow (Result\ content\ type).$$

To demonstrate the meta-path model, we will model an IoTSE instance that queries for "homepages of IoT-enabled light bulbs which are reporting an abnormal energy consumption" as an example. This query can be decomposed into two subqueries: "finding the virtual representative of things, which are light bulbs" and "finding the virtual representative of things, which are reporting an abnormal energy consumption". The first subquery involves assessing each discovered representative (Representative) to determine whether it belongs to a light bulb (Thing) and returning the representative of that light bulb (Representative) as the search result. This subquery can be modelled with the meta-path $R \rightarrow T \rightarrow R$.

The second subquery involves assessing each discovered sensing data stream (Dynamic IoT content) to detect an abnormality, finding the Thing that offers such data stream, and returning the representative of that thing (Representative) as the search result. This subquery can be modelled with the meta-path $D \rightarrow T \rightarrow R$.

By aggregating the two subqueries, we can model the IoTSE instance with the aggregated meta-path $R + D \rightarrow T \rightarrow R$. The IoTSE class addressing this meta-path is the second most common class in the IoTSE literature. The following section on a meta-path-based classification system for IoTSE will discuss more details.

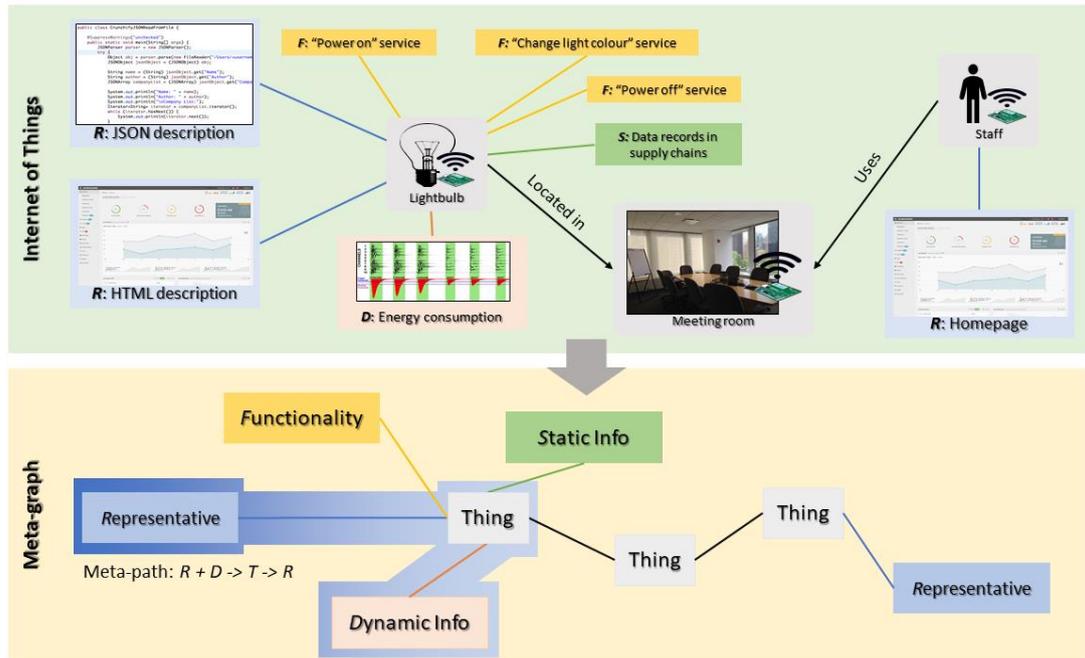

*Figure 4. (Upper) Meta-graph from a concrete IoT network, and meta-path $R + D \rightarrow T \rightarrow D$.*

# 4  A Meta-path-based Classification System for IoTSE

IoTSE instances can be classified in various dimensions, from implementation technologies [14] to query processing behavior [15, 16] and the maturity of their development [17]. Alternatively, we can classify IoTSE instances by their meta-paths, which provide succinct and comprehensive description of their functionality via the type of queries that they support. As mentioned previously, the form of a query that an IoTSE instance addresses influences its internal operations and, therefore, determines its solution space. A meta-path-based classification system will provide insights on what an IoTSE instance is and what it should do, according to the IoTSE literature.

We modelled the IoTSE prototypes selected in Section 2 with the meta-path model and identified 8 types of meta-path (Figure 5). Each meta-path represents a class of IoTSE.

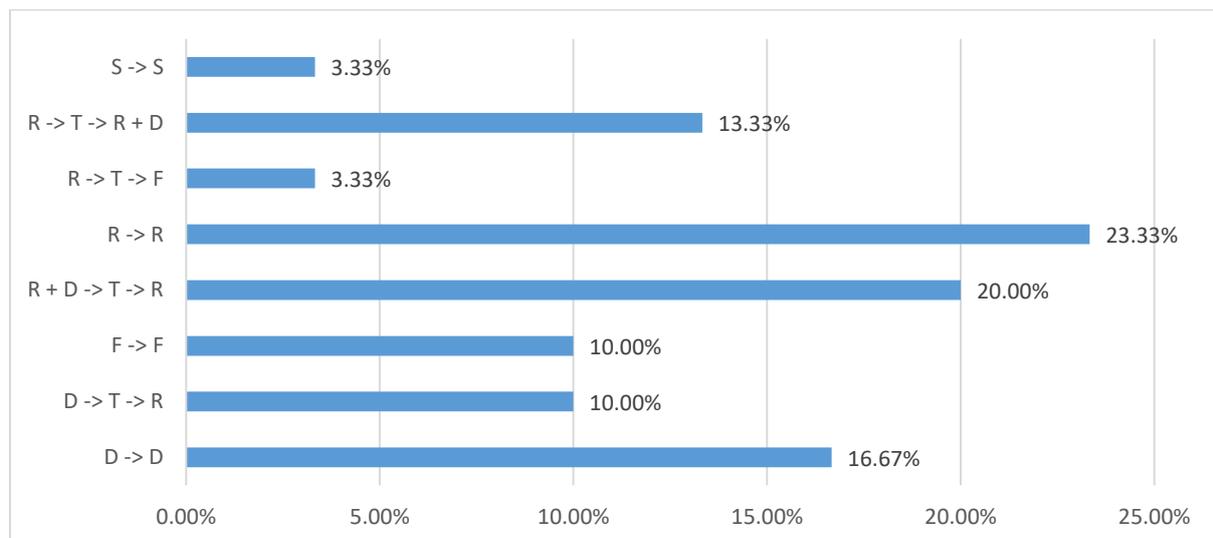

*Figure 5. Distribution of meta-path types among reviewed IoTSE works.*

## 4.1 $R \rightarrow R$ Class IoTSE

This IoTSE class is the most popular in the literature. Instances of this class resolve queries on ID, metadata, or content of representative IoT content, and return matching representatives as search results. The popularity of this class is a surprising finding, as it does not utilize distinctive content types of IoT, such as sensing data and actuating services, nor the relationship between IoT contents and things.

ForwarDS-IoT [18] is an IoTSE prototype processing queries on semantic description of things, stored in a federation of repositories. Queries in ForwarDS-IoT specify conditions on metadata of IoT things and are translated into SPARQL queries. This prototype supports both synchronous and asynchronous query processing. DiscoWoT [10] is another prevalent prototype belonging to the $R \rightarrow R$ class. This system accepts identities of IoT contents as "queries" and returns representation of the given contents in a common format as "search results". Its operation is based on crowd-sourced strategies for translating different types of resource description into the common format. Coverage of strategies represents "discovered IoT content" of DiscoWoT.

## 4.2 $D + R \rightarrow T \rightarrow R$ Class IoTSE

This class of IoTSE resolves queries on both sensing data streams (i.e., dynamic IoT content) and representatives of IoT things, and returns the representatives of things that satisfy both query criteria. Efficient processing of dynamic IoT content is a primary challenge of this IoTSE class. The following two works represent two prevalent approaches to address this challenge.

IoT-SVK [11] searches for IoT things based on their textual description and real-time sensing values, with respect to spatial and temporal constraints. This search engine collects sensor readings continuously in parallel to the query assessment. To handle the constant influx of sensing data, IoT-SVK utilizes two B+ trees and an R tree indexes, which are distributed across a hierarchy of indexing servers. Dyser search engine [7] also searches for IoT things based on their description and real-world states, which are derived from their real-time sensing values. Different from IoT-SVK, Dyser does not collect IoT content. Instead, it contacts things to validate their states for every query. To improve the efficiency of this operation, it predicts sensing values by assuming the existence of repeating periods in sensing data and ranks things according to this prediction to minimize the number of things to validate.

## 4.3 $D \rightarrow D$ Class IoTSE

Search engine instances of this class process queries on metadata and content of sensing data streams and return relevant streams as search results. Key distinction of search engines in this class from the previous one is their focus on low-level sensor readings, instead of high-level states derived from readings.

CASSARAM [19] queries sensing data streams on their contextual information, such as availability, accuracy, reliability and response time. It is motivated by the lack of the search functionality for an increasing number of sensors with overlapping capabilities deployed around the world. CASSARAM utilizes an extension of the Semantic Sensor Network Ontology (SSNO) [20] to describe the contextual information. A user would query this ontology with a SPARQL query generated by the graphical user interface of CASSARAM. This interface also captures the references of the search user. The Euclidian distance between matched sensors and the user reference in a multidimensional space built from different types of sensor contextual information is used for ranking purpose. Top ranked sensing streams are returned as search results.

## 4.4 $R \rightarrow T \rightarrow R + D$ Class IoTSE

This class of IoTSE can be considered as an extension of the class (R -> R). Instances of this class resolve queries on ID, metadata, or content of representative IoT content, and return both representatives and sensing streams of matching things as search results.

Snoogle [9] is a representative prototype of this IoTSE class. It resolves queries on the textual data stored in IoT things to identify and locate the relevant things. Essentially, Snoogle is a text retrieval system operating on distributed, low-powered repositories. It utilizes a distributed top-k query algorithm with pruning, based on the characteristic of flash memory and Bloom filter, to increase the efficiency of the operation. The representative of matching things, along with their location at the query time (i.e., dynamic information IoT content), are returned as search results.

## 4.5 $D \rightarrow T \rightarrow R$ Class IoTSE

This class of IoTSE resolves queries on various aspects of sensing data streams and returns the digital representative of things possessing the matching sensing data streams. Different from class $D + R \rightarrow T \rightarrow R$, search engines in this class do not consider other features of things.

Content-based Sensor Search (CSS) [21] is a representative prototype of this class. It searches for IoT-enabled sensors that produce measurements within a certain range for a certain time prior. CSS contacts sensors to validate their values during query processing instead of collecting IoT content a priori. It utilizes time-independent prediction models (TIPM) to rank sensors based on their probability of having the queried state. These models assume that a sensor reading which is frequently and continuously reported by a sensor in the past has a higher probability to be its current reading. The details of the sensor nodes providing the matching streams are returned as search results.

## 4.6 $F \rightarrow F$ Class IoTSE

This class of IoTSE resolves queries on functionality of IoT things. Considering the popularity of functionality content in real world usage scenarios of IoT (e.g., smart home), the limited support for this IoTSE class is a surprising finding. The lack of public datasets and standards for functionality content might have contributed to this limitation.

Mrissa, et al., [22] present a search and discovery mechanism for functionalities of physical entities. It aims to discover and expose high-level functionalities of a physical entity that can be realized by a combination of its low-level physical capabilities and functionalities exposed by other entities in the immediate area. These functionalities and capabilities are described in a shared ontology. Each physical entity queries this ontology with a set of SPARQL queries encapsulated in Java functions. This work is part of the avatar architecture from the ASAWoO project[2].

## 4.7 $R \rightarrow T \rightarrow F$ Class IoTSE

This class of IoTSE resolves queries on the representatives to find relevant things and returns functionality of those things as search results. Kamilaris, et al., [23] propose an IoTSE instance that utilizes a DNS-like mechanism to search for IoT things and return their functionalities. These functionalities are presented as RESTful Web services and capable of self-describing with specifications written in the Web Application Description Language (WADL).

---

[2] https://liris.cnrs.fr/asawoo/doku.php

## 4.8 $S \rightarrow S$ Class IoTSE

This class of IoTSE resolves queries on static information IoT content. Matching content is returned as search results. Microsearch [24] is an instance of this IoTSE class. It is essentially a down-scaled information retrieval system operating on sensor nodes with very limited computing and storage resources. It indexes small textual documents stored in the sensor node and returns the top-k documents that are most relevant to the query terms given by a search user.

# 5 Open Issues

As IoTSE research and engineering is a complex and relatively new area, researchers and practitioners face several types of technical challenges. In this section, we discuss four open issues, derived from the existing literature, that affect most classes of IoTSE.

## 5.1 Building datasets for IoTSE research

Large-scale, open datasets that contain IoT content, sample queries, and ground truth, are critical to IoTSE research. They negate the need for difficult-to-replicate experiments and simplify the experimentation and evaluation of the research works on IoTSE. Research works on IoTSE have utilized some sensing datasets, such as Intel Lab[3], NOAA[4], bicycle rental[5], taxi GPS[6]. Actuating functionality datasets, on the other hand, have not been found in the existing literature. Availability of the sample queries has also been limited, as they tend to be private property of industrial IoTSE instances [6]. Providing access to IoT datasets is a challenge due to their massive size, reaching 21 Terabytes a day [6], and their potential threats to privacy. Given these opportunities and challenges, building open IoT datasets is essential in the IoTSE research.

## 5.2 Ranking IoT contents by their natural order

Natural order ranking denotes the ordering of content by their intrinsic characteristics instead of their relevance to a given query. In large data collections where a massive number of data items can be relevant to a query, a search engine must rely on natural order ranking mechanisms to order and deliver the most relevant search results to its query clients. For example, the ranking of Web pages based on their importance by using link analysis algorithms such as PageRank is a form of natural order ranking.

As the anticipated size of the IoT is even more extensive than the Web, we anticipate that the natural order ranking mechanisms for the IoT content will be an exciting and challenging research topic that will play a crucial role in IoTSE. The first problem in this topic would be defining a natural order that is applicable across different IoT content. For the Web, the level of authority is the natural order of Web pages. For the IoT, what would be the natural order of the heterogeneous IoT content? When this natural order has been defined, the next problem would be developing mechanisms to calculate it on the IoT-scale.

A potential solution to natural order ranking could rely on the quality-of-service metrics of IoT services. Another potential approach could reuse the solution of the Web by constructing a network of hidden links between IoT things [13, 27] and applying link analysis algorithms such as Page Rank and its variants to devise a natural ordering of content in the IoT.

---

[3] http://db.csail.mit.edu/labdata/labdata.html
[4] https://data.noaa.gov/datasetsearch/
[5] https://www.bicing.cat/
[6] https://github.com/roryhr/taxi-trajectories

## 5.3 Security, Privacy, Trust

IoTSE instances have the potential to detect and retrieve anything in the IoT, at any place and any time. They bring a wide range of benefits to human users and software agents but also present significant security and privacy risks. IoTSE instances can track a person, monitor an area without consent [25], and spy into warehouses of competing businesses [17]. Perpetrators can also take advantage of IoTSE to propagate malicious sensing information and actuating services. As future IoT applications might rely solely on IoTSE to acquire IoT content for their operation, misleading information propagated by IoTSE can have severe impacts. For example, by planting sensors that imply a restaurant is full, competitors can drive it out of business. Addressing security, privacy, and trust issues, therefore, is arguably more critical to the success and adoption of IoTSE compared to perfecting its discovery and search algorithms.

## 5.4 Facilitating composition and reuse of IoTSE solution

Across different classes of IoTSE, we have observed shared internal operations such as content discovery, indexing, and searching, albeit with different implementation to serve different types of IoT content. We have also observed the overlaps between various meta-paths, such as between [D + R -> T -> R] and [D -> T -> R]. These observations suggest that prior IoTSE instances can be reused to improve other instances or compose new instances, which might utilize a different meta-path.

Realizing composition and reuse of IoTSE solutions require a common IoTSE architecture and a supporting software infrastructure to support the development, accumulation of IoTSE components, and the engineering of IoTSE instances from those components. Tran, et al. [26] propose to utilize a shared software library to facilitate the development of reusable, composable IoTSE components and support the composition of these components into operational IoTSE instances. The approach could be improved by reducing the constraints of the shared library on component developers and simplifying the distribution of components in an IoTSE instance. The Service-oriented Architecture (SOA) is a potential solution to this problem, due to its enforced separation of concern between services and its native support for composition.

## 6 Conclusion

Internet of Things Search Engine denotes a software system responsible for discovering and resolving queries on contents of the Internet of Things. Due to the diversity of IoT contents, developing IoTSE is a complex and diverse problem that is still relatively immature. This article introduces concepts, models, and a classification system for IoTSE, which have been generated from a structured and comprehensive study of the literature on IoTSE. We have categorized the latest works into eight classes of IoTSE and presented four major open issues that impact all classes of IoTSE.